\documentstyle[12pt,epsf]{article}                        
\begin{document}
%Para dar tamagno a 12pt%%%Si quieres 11pt pon 11pt%%%%%%%
\font\blackboard=msbm10 at 12pt
%%%%%%%%%%%%%%%%%%%%%%%%%%%%%%%
 \font\blackboards=msbm7
 \font\blackboardss=msbm5
 \newfam\black
 \textfont\black=\blackboard
 \scriptfont\black=\blackboards
 \scriptscriptfont\black=\blackboardss
 \def\bb#1{{\fam\black\relax#1}}

\baselineskip18pt

\newcommand{\r}{{\bb R}}

%\baselinestretch{1.5}

\thispagestyle{empty}

\begin{flushright}
\begin{tabular}{l}
FFUOV-98/08\\
{\tt hep-th/9810060}\\
%\today
\end{tabular}
\end{flushright}

\vspace*{2cm}

{\vbox{\centerline{{\Large{\bf The Principle of Equivalence as a Guide
}}}}}

\vspace{0.8cm}

{\vbox{\centerline{{\Large{\bf
towards Matrix Theory Compactifications.}}}}}

\vskip30pt

\centerline{Jes\'{u}s Puente Pe\~{n}alba
\footnote{E-mail address:
  jesus@string1.ciencias.uniovi.es}}

\vskip6pt
\centerline{{\it Dpto. de F\'{\i}sica, Universidad de Oviedo}}
\centerline{{\it Avda. Calvo Sotelo 18}}
\centerline{{\it E-33007 Oviedo, Asturias, Spain}}

\vskip .5in

\begin{center}
{\bf Abstract}
\end{center}

%\vskip 18 pt

   The principle of equivalence is translated into the language of the
world-volume field theories that define matrix and string theories. This
idea leads to explore possible matrix descriptions of M-theory
compactifications. An interesting case is the relationship between $D=6$
${\cal N}=1$ $U(N)$ SYM and Matrix Theory on $K3$.

\newpage

\section{Introduction.}

   In the last few years, important changes in the picture we have of the
physics beyond the Planck scale have occurred. The publication of several
articles that proposed and gave evidences for new dualities has helped to
fill even more the space of theories M-theory has become. It was the case of
Matrix Theory \cite{bfss,intro,dlcq} and the AdS/CFT conjecture \cite{malda}.

   The guide for all has been the use of different quantum field theories to
describe systems that include gravity. Based on D-brane world-volume
physics, different limits have been found in which particular configurations
of D-branes in certain backgrounds were able to account for all the degrees
of freedom of the system.

   What is most surprising of all this is that a wide variety of field
theories seems to contain, at least in some limit, such an abstract
symmetry as Einstein's principle of equivalence. The purpose of this article
is to show where it is hidden among the properties of the field theories and
to find out that, in fact, it holds for all scales.

   Once that is done, one can write down some minimum properties that allow
to construct a theory of quantum gravity and take advantage of them in order
to find some examples apart from the ones that are already known.

   Led by this, I show the good features of one interesting case: $D=6$
${\cal N}=1$ $U(N)$ SYM. I discuss its possibilities and how it could match
into the moduli space of M-theory.

\section{Quantum field theories and the principle of equivalence.}

   All solutions that have been tried until today to the problem of gravity
quantization consist of a world-volume interpretation of one (Matrix Model)
or two (String Theories) dimensional quantum field theories. It is an
interesting task to find out which are the common characteristics that allow
all these theories to describe gravity and see how those properties
constrain the construction of other possible theories.

   One may think, to begin with, that the only requirements for the theories
are to be quantum and to include in some way the principle of equivalence.
This is achieved if the theory holds the following conditions:

\bigskip

  - To be a quantum field theory. In principle, the number of dimensions is
not important.

  - To be renormalizable, because we want the theory to be consistent. I
shall write more about this topic later.

  - To have a sufficiently complex vacuum so as to be described by
some continuous set of parameters that define a manifold of the kind of
$\r^d$ or any other that could be interpreted as a target space. If there
are fermionic (Grassmann) parameters, they should be able to be arranged
into spinnorial representations of the rotation group of the bosonic space.
This is in order to have a target super-space formulation. It is essential
that all the parameters that define the vacuum have a (super)geometrical
interpretation.

  - This way, the theory can describe the dynamics of one single object. If
we want a theory with a full second quantization, we had better find a
family of quantum field theories whose moduli spaces are $(\r^d)^N$, for
every possible $N$. In any other case, the world-volume will be unique, and
multiple objects may only be included in a more indirect way. In string
theories, they appear as different boundaries (vertex operators) of the
world-sheet.

  - The fields whose vacuum expectation values define the target manifold
must possess a `flavour' symmetry with a group $SO(d)$ or $SO(d-1,1)$.

\bigskip

   These conditions are the world-volume implementation of the principle of
equivalence. This is understood noting that, up to first order in
perturbation theory (that means: in a small enough neighbourhood of the
observer where it is not much affected by interactions), all physical
magnitudes depend exclusively on the positions of the objects (vacuum
expectation values of scalars), because they are the only quantum numbers
that define the vacuum. In particular, we can consider a probe -an observer-
surrounded by some nearby objects; as they stand approximately in the same
position, all of them suffer the same acceleration produced by the
interaction with the rest of the universe. In this situation, the observer
defines an inertial frame that does not feel at all the action of gravity
over a sufficiently small volume around him. Moreover, the `laws of nature'
that the observer measures respect an
$SO(d-1,1)$ symmetry, that is, the kinetics is that of special relativity.
If the symmetry group is $SO(d)$, instead, we can either keep the theory as
Galilean or try to interpret it as a light-cone formulation of a
$d+1$-dimensional theory.

   If we act like this we can recognize all the interactions as
gravitational and therefore we can always find a geometrical interpretation
of the involved forces. This holds independently of how small we take the
typical scale to be.

   There is a subtlety regarding the inclusion of fermions. They induce
supersymmetry in the target space. In general, if the supermultiplet is
large enough, this could add interactions mediated by vector or scalar
fields with spin one that may include gauge groups. The argument above must,
therefore, be generalized because interactions may depend not only on the
bosonic coordinates, but also on the fermionic ones. That means that the
interactions will be supergravitational. In this case the principle of
equivalence is enhanced so that the local symmetry observed in the inertial
frame is not only the one spanned by the Poincar\'e group but the whole
supersymmetry algebra.

   Somehow, this conditions represent the minimum properties a world-volume
theory must have in order for the target theory to include gravity.

\section{Known examples: String Theories and the Matrix Model.}

   Let us first see what is the most simple example of the philosophy
outlined in the previous section: String Theories. Polyakov and
Green-Schwarz actions are just the actions of two-dimensional free scalar
fields. The vacuum of the theory is parameterized by the vacuum expectation
values of the positions of the strings, that, indeed, define a Minkowskian
space with $SO(d-1,1)$ symmetry. The theory does not have a complete second
quantization because it is not a family, but just a single field theory. The
only way in which several strings can be included is by performing a number
of punctures in the single world-sheet when scattering amplitudes are
calculated. So multiple strings are obtained taking advantage of the fact
that the world-sheet has two dimensions and so an arbitrary number of
boundaries (holes) can be put on it.

  Let us now move on to the Matrix Model. Exactly the same arguments can be
made to see that all conditions are held \footnote{There are two papers
\cite{sphere} that explicitly say that finite-N Matrix Model breaks the
principle of equivalence even for long distances. This was deduced assuming
the existence of backgrounds with vacuum expectation values for off-diagonal
terms. Those are just quantum fluctuations. The statements I make above only
refer to stable vacuum configurations of the theory.}. This ensures that the
principle of equivalence is an exact symmetry for all scales, which is not a
trivial result. There are certain subtleties related in part to the
light-cone interpretation that I want to clarify. In principle, the
conditions described in the first section guarantee that we are describing
objects of some class moving in a gravitational background, but that does
not mean that the objects naturally come out of the gravity theory. To make
it clearer: it is not necessary that the graviton should be one of the
objects that the theory can explicitly describe.

  This can be better understood with an example. Take the Dirac-Born-Infeld
gauge theory that describes the short-distance dynamics of general
D-$p$-branes with positive Ramond-Ramond charge. Regardless of whether or
not there is any realistic physical situation in which the only significant
degrees of freedom are those of the D-$p$-branes with positive charge, the
action includes, among others, the gravitational interaction. What I am
trying to say is that the conditions in the second section do not impose
the action to be realistic nor to completely determine a whole theory, just
to describe gravity in a quantum manner.

  As it is clear that gravitons are well defined asymptotic states in
nature, we can only interpret D-brane actions, and some others that I shall
later consider, as partial descriptions or formulations adapted to particular
physical configurations where, for example, graviton states are irrelevant.

   We begin now a search, among some simple classes of theories, for some
examples, apart from those already known, that could describe interesting
physics.

\section{Bosonic theories.}

   One should now see how other gauge theories can be interpreted as
gravitational ones. I shall begin with bosonic Yang-Mills theories in
different dimensions. It is necessary to dimensionally reduce the action, at
least in some directions because the vacuum expectation values of gauge
fields are pure gauge and do not span any moduli space. The dimensional
reduction gives us a number of scalars that determine the dimension of the
configuration space.

   The r\^ole of the gauge group is to define the vacuum and to produce a
particular potential. The $U(N)$ set of groups is particularly well adapted
to include a many-particle configuration space into its vacuum degeneracy,
so we shall study it first.

   When one makes a field theory calculation, similar to that in \cite{bb}
the long distance potential that is obtained is $V \propto r^{1+p}$
independently of the original dimensions of the YM. $p$ is the number of
spatial dimensions that have not been reduced. This potentials only coincide
with a Newton long-distance limit in two dimensions and therefore, in the
general case they would represent gravitational fields that do not vanish in
the infinite. They are not of any physical interest.

   We only have left the possibility to interpret that the objects we are
describing are not point particles, not even branes, but extended objects
like bodies in three dimensions with a particular mass distribution.
Coordinates should, in this case, parameterize the center of mass of each
object. If we choose the correct distribution, we can classically obtain
virtually any potential that grows with the distance faster than the
Newtonian. Except, maybe, for very particular situations, this
interpretation is hardly realistic.

   We do not need, however, to force the theories to describe multi-particle
systems. Take, for instance, the $SU(2)$ theory. The vacuum defines a single
particle geometric space. We can now interpret the theory as the description
of the dynamics of one particle moving in a particular background which is
responsible for the potential. The theories do not seem to be very useful
although it should be pointed out that $V \propto r$ potentials appear over
thin shells around spherically symmetric objects in any dimension. It is not
clear, however, what kind of boundary conditions should be imposed.

   Without any further investigation on this topic, we move on to some, more
interesting theories.

\section{Supersymmetric theories.}  

   Supersymmetry can be added introducing fermions in the world-volume
theory. The easiest way to do this is by imposing supersymmetry among the
world-volume fields. This can only be done if the number of fermionic
degrees of freedom that we get can be arranged into a spinor of the target
space. There, they will be interpreted as fermionic coordinates
(supercharges). This happens in 10, 6 and 4 dimensions. There are probably
more cases in higher dimensions, but with little physical interest.

   For them to be fundamental theories -by that I mean: defined in a whole
phase space, it is necessary that they should give appropriate long range
forces, that is, Newton potentials. If the potential is stronger than that
-it was the case in the bosonic theories- they cannot be used to describe
open spaces, because the graviton fields do not vanish in the infinite. If
the potential is weaker, it would at most describe a gravitational theory
whose action does not contain the scalar $R$, but higher powers or
derivatives of it. To my knowledge, these do not have any use in any
realistic situation.
  
   It is possible to make the calculation of the potential at tree level. In
ten dimensions it was made in \cite{bb,rutgers}, and the other cases are
simple extensions. Again, we consider these three theories reduced to $0+1$
dimensions. The gauge group is $U(2)$ because we need the potential between
two particles. The results of these computations are
\begin{eqnarray}
V_{D=10}(r) &\propto& \frac{v^4}{r^7} \nonumber \\
V_{D=6}(r) &\propto& \frac{v^2}{r^3}\hspace{1cm}  {\mbox{and}} \nonumber \\
V_{D=4}(r) &\propto& \frac{v^2}{r^3}.
\end{eqnarray}

  The former case can be recognized as the potential between D0-branes, and
the resulting fundamental theory is, of course, Matrix Theory. It is also
very interesting that the six-dimensional case also gives a Newton potential
and, therefore, a long-distance metric that reproduces its effects can be
written. It is

\begin{equation}
ds^2=\left(1+\frac{g_{YM}}{2 r^3}\right)\left(dr^2+r^2 d\Omega_4^2\right).
\end{equation}

\section{$D=6$  ${\cal N}=1$ $U(N)$ SYM.}

   In our more or less systematic search for realistic gravitational
theories, we have eventually found this good candidate. I shall concentrate
on it from now on. It can be extended to a Born-Infeld-Dirac type action
with an $SO(5,1)$ space-time symmetry and, therefore, relativistic in the
target space. It is renormalizable and contains gravity. In principle, it
describes one type of object (BPS saturated) moving in a supergravitational
background. The Lagrangian does not explicitly include free graviton states
so one could question if it is enough to define a whole theory, that is, if
it contains all the information needed. The argument supporting this is that
gravity is different from any other interaction in that it acts exactly the
same over any physical object independently of any other quantum number but
the stress-energy tensor. That means that any object can probe it and
extract the gravitational Lagrangian because all behave the same. Analyzing
the potential or the force between the particles in terms of their positions
and velocities, it is possible to identify a complete action formed by an
infinite series of terms in derivatives and powers of the curvature scalar
$R$. This is always possible because what we have is a field of
accelerations, rather than a strength field. The sixteen supercharges that
we know the theory to have help us complete every term with the necessary
fields. To make it short, what we should do is calculate scattering
processes, then deduce the potential terms created by them and finally,
identify the extensions of supergravity that reproduce them. I have already
taken the first step when I have compared the supergravity `Newton'
potential to the SYM one-loop potential. This is exactly the philosophy
followed by Matrix Theory. The main conjecture there is to assume that all
the objects that appear in the theory are either gravitons or bound states
of them. In particular, the `algorithm' I have just described to obtain the
supergravity series has been partially applied to the D0-brane action to get
the $D=10$ ${ \cal N}=2$ supergravity action and some corrections
\cite{ricci}. In fact, as Matrix Theory compactifications and simple D-brane
interactions have shown \cite{bfss,pol}, any D-brane action can be used to
calculate gravitational interactions if the momentum transfer is restricted
to the directions which are orthogonal to the D-planes. The difference
between using the complete DBI action and just the SYM term is that by
taking the light-cone in the eleventh direction (using SYM), it is made sure
-according to the conjecture- that all the objects are described by the same
exact action. That is, in the fundamental string action, D-branes appear as
non-perturbative, solitonic states; the D-brane action is related to it
through $S$ and $T$ dualities so it contains the same information, but
differently organized so that it is useful in different physical
circumstances. The BFSS's guess is that the Matrix Model Hamiltonian
perturbatively describes all the objects of M-theory in the light-cone.

   In the six-dimensional case that we are studying, we do not have so many
dualities to guarantee the completeness of the theory but the fact that we
can describe at least some quantum objects interacting through gravity up to
any energy or length scale is very encouraging.

   One may think that the lack of so much supersymmetry as in the Matrix
Model, and therefore, the lack of some non-renormalization theorems may
cause some problems in the relation between the SYM and the supergravity.
The answer in no. The only requirement is that the theory be renormalizable
because the comparison is made just by calculating physical magnitudes. In
particular, the potential is obtained studying scattering processes. The
fact that the coefficients of the potential are or not corrected by higher
loop terms is not important because we are not comparing the theories loop
by loop, but their physical effects as a whole. That is, the whole SYM
series is equivalent to the whole -infinite- supergravity series.

   This is all we can say about this theory, considered independently of M
and string theories. It seems to be complete, in the sense that it possesses
all the information of a six-dimensional extended supergravity theory, valid
up to all scales. However, the description of many degrees of freedom
(including gravitons) gets very complex at small distances because one has
to construct the supergravity series that gives more and more complicated
interactions among the gravitons as one considers smaller distances. In
fact, the mechanism that would regularize the theory at high energies
-parallel to the appearance of very degenerated massive states in string
theory- is the presence of massive bound states of gravitons. This is the
same guess as in Matrix Theory, where the bound states include, among
others, membranes and, therefore, strings in ten dimensions. This makes it
very difficult to know which are the relevant degrees of freedom at high
energies or find, for instance, a partition function for the complete
theory.

   T-duality shows that whenever a small volume limit is taken in any string
or matrix theory, states with discrete momentum different from zero are, in
effect, decoupled as in any other field theory, but there always exists a
set of other quantum numbers (string windings in the simplest cases) that
gather into a continuous spectrum that can be interpreted, thanks to
T-duality, as the opening of new dimensions in the dual theory. This does
not seem to be the case: there is not any obvious Kaluza-Klein spectrum and
the continuous spectrum is just six-dimensional. This gives us three
possibilities: the theory might be disconnected from M-theory
compactifications; it could be connected, but with a K-K spectrum that is
hidden as some bound states of gravitons; finally the theory may need to be
completed with more fields (or simply more information). The difficulties in
calculating bound states in supergravity series seems to make it impossible
to decide which is true, but I shall give some hints.

\section{A possible relation to the Matrix Model.}

   Looking at its form, very similar to the usual Matrix Model, one is
pushed to try to locate this six dimensional model in a `nearby place' in
the moduli space of M-theory. I shall now study this possibility. 

   The Lagrangian describes BPS states whose supermultiplet is generated by
8 supercharges. This means that the effective theory should have 16 of them,
and should not be chiral. In six dimensions, there are several theories
whose low energy supergravities share those characteristics, but all of them
are related by different dualities. The most interesting to us for its
simple relation to M-theory is type IIA string theory compactified on a $K3$
manifold. The $K3$ may be taken to any singular limit with -for example-
orbifold singularities. This changes the degeneracy of the effective model
as well as the gauge symmetry. The guides that will lead us to find out the
particular situation that may be described by $D=6$ ${\cal N}=1$ $U(N)$ SYM
are its spectrum of objects and their behaviour.

   The only objects that are explicitly present are heavy point-like BPS
states. Just like Matrix Theory, it contains information about other
massless objects like six-dimensional gravitons but they are only included
in the interactions.

   The only way to simplify the spectrum of M-Theory so as to have just one
kind of object seems to be taking the light-cone in the eleventh dimension,
that is, to use the Matrix description. So we had better look for the Matrix
model of type IIA string on a $K3$. This has already been studied
\cite{compact,five} and seen to be described by the theory of longitudinal
M-5-branes wrapping $S^1\times K3$ or equivalently IIA D4-D0-brane bound
states just wrapping the $K3$. The particular field theory is not explicitly
known, but some properties have been studied.
 
   Indeed, $D=6$ ${\cal N}=1$ $U(N)$ SYM does have enough supercharges to generate
the hypermultiplet of the five-brane. Moreover, wrapped five-branes behave
like point-like objects in the open directions. The straightforward guess is
that the diagonal elements of the YM scalar matrices should be interpreted as
the transverse positions of the five-branes. 

   However, we only have one arbitrary parameter to fix -the coupling, so it
is clear that we cannot reproduce the moduli space of the $K3$. To solve
this, we have to see what happens to the link between them: the
supergravity. Six-dimensional type IIA supergravity is determined by the
string coupling and the vacuum expectation values of the scalars that appear
after the compactification. The latter can be arranged into a $24
\times 24$ matrix that can, at generic points, be diagonalized to give,
basically, the $24$ couplings of a gauge group that is broken to
$U(1)^{24}$. Therefore, the moduli space is, generically, 25-dimensional.
Our case is, however, simpler, because we are describing just one kind of
object. It is a BPS state and so its gauge charge is the same as its mass.
That is why, the interaction of our object is determined by one coupling. In
general, in the Matrix models, the complexity of the moduli space is carried
by the manifold in which the field theory of the model is defined over.

   The only physical situation where the particular form of the manifold
looses importance is when its size is reduced to zero. In the Matrix models,
the field theories are defined over the T-dual manifold. When none of the
manifolds -neither the original nor the dual- is large or small, the target
moduli space is defined by the vacuum expectation values of the scalars and
the Wilson lines of the gauge fields. If the T-dual volume is vanishingly
small, the field theory is defined in a point and we can therefore expect a
quantum mechanical formulation. Besides, when the volume is small, all the
internal degrees of freedom can be integrated out so that the only physical
magnitudes that survive -and therefore should appear in the action- should
be the positions of the particles. This is a kind of `dimensional reduction'
of the Matrix model. This argument is also valid for large volumes because
of the self-T-duality of type IIA superstring in six dimensions.

   This tells us that the provisional conjecture that should be proposed is
that the Matrix model that describes the $D=6$ non-chiral ${\cal N}=2$
string theories in the DKPS limit and with $V_{\tilde{K3}} \rightarrow 0$,
is $D=6$ ${\cal N}=1$ $U(N)$ SYM reduced to $0+1$ dimensions. The relation
between the Yang-Mills and the IIA string coupling should be \cite{compact}:
\begin{equation}
g_{YM}=g_S^{-1/3} V_{11}^{-1}
\end{equation}

   This helps us know what kind of limit we should take the volume to. The
relation between the original volume and the T-dual is

\begin{equation}
\tilde{V}=\frac{1}{m_p^{10} R_{11}^{5} V(K3\times S^1)}
\end{equation}
where the volumes are measured in units of the eleven-dimensional Planck
length. In units of $\alpha'$, the original volume is 
\begin{equation}
V_{\alpha'}=\frac{V(K3\times S^1)}{R^5 m_p^{10}} g_S^{10/3}
\end{equation}

   What this is telling us is that the limit we are looking for is that with
a T-dual small volume, which means that the original compact volume must be
large compared to the Planck length, while remaining vanishing as $g_s
\rightarrow 0$ in terms of $\alpha'$. Besides, if we want the D4-brane
description to be valid -weakly coupled-, we should take the original volume
to infinity quicker than $g_S^{-1/3}$. Its world-volume theory is not
renormalizable, but that does not matter as long as all its fields are
decoupled. I shall be more concrete in next section. This is the addition
to the DKPS limit that seems to simplify the spectrum enough to allow us to
use a theory with much fewer degrees of freedom.

\section{Compactifications.}

   Apart form the small dual volume limit, there are other reasons why we
have such a simple theory in six dimensions when the equivalent
compactification on a torus is so difficult to define. Of course, we have
presumably decoupled the gauge theory, which carries most of the complexity,
and besides, the supersymmetry breaking reduces a lot the number of degrees
of freedom. In particular, the supermultiplets are short enough not to
-necessarily- include gravitons. A similar argument was pointed out in
\cite{calabi} when discussing compactifications of Matrix Theory on
Calabi-Yau manifolds.

   Among other things, this allows us to compactify further and see what
happens in lower dimensions at least when half the supersymmetry is broken.
We can choose the manifolds to be tori and take advantage of the T-duality
of gauge theories shown by \cite{taylor}, which is general and independent
of the dimension. This way we could define Matrix Theory on $S^1 \times K3
\times T^d$. As gauge theories are renormalizable up to three dimensions, we
can compactify down to $D=2+1$, where the description is in terms of $D=3+1$
${\cal N}=2$ SYM.

   The situation is particularly interesting when compactifying just one
dimension. In that case, one can be more explicit. We take advantage of the
following set of dualities. We begin with type IIA on $K3\times S^1$ and
make a matrix model of it. T-duality leads us to a IIB D5-brane wrapping
$\tilde{K3} \times \tilde{S}^1$. That is the same as another
$\tilde{\cal{M}}$-theory on $\tilde{T}^5/Z_2 \times \tilde{S}^1$ where one
of the directions of the five-torus is light-like. We can choose any of the
eleven dimensions to define a type IIA coupling constant so we make a flip,
similar to the one in \cite{dvv}, and choose the direction of $\tilde{S}^1$.
That one is not light-like, so we get the non-perturbative definition of IIA
string theory on $\tilde{T}^5/Z_2$. We can get the Lagrangian of this theory
in the limit we are interested in led by the calculation in \cite{dvv}.

   Let us start with the Lagrangian of N D0-branes moving on $K3 \times S^1$.
By T-duality, it becomes a six-dimensional SYM defined over a $\tilde{K3}
\times \tilde{S}^1$. This is true at least at low energies. If the theory is
not renormalizable, it should be appropriately completed at high energies.
If we take its volume to be vanishingly small, we can make a dimensional
reduction of the theory. What we are left with are the zero-modes of the
scalars and of the gauge fields. However, in the $K3$, all loops are
contractible and so no Wilson line can be defined and the zero modes of the
gauge fields are pure gauge except for the component in the direction of the
circle. In this moment we interchange the r\^ole of that circle and the one
related to the string coupling. This puts the Lagrangian exactly in the form
of an ${\cal N}=1$ $D=5+1$ SYM reduced to $D=1+1$ and that is precisely what
we expected. The coupling is

\begin{equation}
g_{YM}=m_S^{-4}V_{K3}^{-1}R_5^{-1}R_{11}
\end{equation}

   The interchange of the fifth and the eleventh dimensions gives a IIB-like
S-duality related, like in the ten-dimensional case, to the modular
transformations of the $5-11$ torus.
   The fact that the theory on the $K3$ decouples helps us deal with a field
theory which could, in principle, be non-renormalizable. Being a well
defined compactification of M-theory, we suppose that the world-volume
six-dimensional field theory exists and that is enough for our purposes. On
the other hand, the T-dual theory may be strongly coupled even when the
original one is weakly coupled; however, after the swap of the two circles
is performed, the string coupling is related to the size of the circle we
choose and independent of the original coupling constant.

   It is interesting to note that when one takes the coupling of the $D=1+1$
field theory to zero, one gets something that could be interpreted to be
a six-dimensional Green-Schwarz action in second quantization. One could be
tempted to relate these to the little strings that move through the
world-volume of five-branes. However, our strings interact through gravity
and so they do not have any relationship.

   If that were the whole story, then a lot of evidence would be in favour
of this theory to be one more of the matrix models. However there are a
couple of things that we have not taken into account. We have not solved the
problem of the lack of continuous spectrum even when we have explicitly used
T-duality so we must have forgotten something. It is the existence of open
string instantons related to the axion $B^{\mu \nu}$'s topological
configurations around the $K3$'s two-cycles. I shall be more explicit, the
Fourier-Mukai transform \cite{horioz} that plays the r\^ole of the T-duality
in the $K3$, changes the original D0-brane with a bound state of a D4-brane
wrapping the $K3$ with another D0-brane. The T-duality along the direction
of the second circle transforms the system into a bound state of a D5 and a
D1-brane. The light degrees of freedom are not only the gauge ones, because
the open strings with both ends attached to the D1-brane can have
world-sheets which are surfaces that wrap non-trivially around the
two-cycles of the $K3$. These states are instantons because they are
localized solutions both in space and time. Their energy is proportional to
the area of the two-cycles, which goes to zero when the volume of the $K3$
is taken to zero. These light states are the ones that provide the
continuous spectrum that we missed. These states are also involved when one
tries to see which is the effect of the T-duality in the S-dual theory, the
heterotic string on a four-torus.

   This means that the SYM quantum mechanics cannot, by itself, describe all
the theory. We must add the instantons. The final conjecture should therefore
be that: `M-theory on $S^1 \times K3$ in the light-cone is dual to the
system formed by the $D=6$ ${\cal N}=1$ $U(N)$ SYMQM BPS states in
interaction with the instantons'. The theory is not as simplified as we
expected. The situation is somewhat similar to what happens in the matrix
model for the six-torus. There, there are two objects that do not decouple:
the D6-branes and the D0-branes \cite{compact,calabi}. That case is,
however, worse because the light D0-branes contain gravity multiplets. In
our case, the instantons have the same states as open strings at tree level
-that is precisely what they are- so their description is much simpler. In
fact, the inclusion of instantons just means the inclusion of open strings
attached to the 0-branes in such a way that their world-sheet is a sphere
around a two-cycle of the $K3$. This makes their dynamics relatively easy to
treat.

\section{Degeneracy of the quantum states: gauge symmetry and bound states.}

   One characteristic I have not written about yet is the gauge group. In
M-theory, it is given by the Kaluza-Klein interpretation of the higher
dimensions. In the case we are dealing with, we can take advantage of the
S-duality we have at hand between type IIA string theory on $K3$ and the
$E_8 \times E_8$ heterotic string on a four-torus. The gauge group at a
generic point in the moduli space is $U(1)^{24}$, but it can be enhanced at
singular points where some degrees of freedom become massless. From the type
IIA point of view, that happens whenever the $K3$ is deformed to develop a
singularity and some of its two-cycles are reduced to zero size. In the
perturbative theory, the D2-branes that wrap the degenerate two-cycles
appear as massless point-particles in the six-dimensional theory and are the
ones responsible for the symmetry enhancement.

   It is straightforward to understand that from our non-perturbative point
of view, the instantons, whose energies are proportional to the surface of
the two-cycles, will take the r\^ole of the D2-branes. That is, the
particular form of the $K3$, including its singularities, is encoded in the
dynamics of the instantons. They reproduce the whole moduli space and
generate the different target gauge groups. 

   When the gauge group is non-abelian, the 0-branes acquire another index
that works as a quantum number that does not appear in the Hamiltonian. The
freedom to add numbers to the labels of the states always exists in quantum
mechanics. In the same way that Pauli spin is added by hand, any other
number could also be added. The justification must come from a more general
theory. In our case, the theory in eleven dimensions.

   There is an important feature that this theory should have in order to
relate it to a matrix model of M-theory: bound states at threshold. They
represent the Kaluza-Klein partons that come out from the compactified
eleventh dimension. The calculation of the Witten index of the SYM was
performed in \cite{yi}. The result was that there were no such states. This
is what we expected because part of the interaction that we have in our case
was not considered. The possibility is not negligible that even the fields
that we have decoupled in the small-volume limit may have some `topological'
effects in the Witten index calculation, which is basically a problem of
field counting (I mean: one field with nearly infinite mass is still one
field).

   This does not prevent us from dealing with the bound states. We know
from their existence, thanks to the dualities and the limits we have taken,
and that will be enough for dynamical purposes. The only change is the same
as in the gauge degrees of freedom: we should add another index that tells
us that whenever two 0-branes coincide in the same point, the degeneracy is
two (there is one free and one bound state).

\section{Conclusions.}

   I have shown how the principle of equivalence is hidden in the
world-volume interpretation of quantum fields. The properties of the fields
in the parametric space of the trajectory have little in common with their
characteristics in the target space. In particular, most symmetries of the
target theory come from symmetries of the vacuum configurations of the
world-volume fields. I have studied how this is achieved in Matrix and
String theories.

   This insight leads to uncover certain theories that can probably be
useful to study Matrix Theory compactifications. $D=6$ ${\cal N}=1$ SYM has
been studied with more depth and seems to be the correct theory that,
together with the dynamics of some open string instantons, is able to
describe M-theory on $S^1$(light-like)$\times K3$.

\section*{Acknowledgements.}

   I am grateful to M. A. R. Osorio and M. Laucelli Meana for useful
discussions.

\end{document}